# ENTROPY BASED DETECTION AND BEHAVIORAL ANALYSIS OF HYBRID COVERT CHANNEL IN SECURED COMMUNICATION

Anjan K[1], Srinath N K[1] and Jibi Abraham[2]

[1]Department of Computer Science and Engineering,
R V College of Engineering, Bengaluru, India
[2]Department of Computer Engineering and Information Technology,
College of Engineering, Pune, India

*ABSTRACT*

*Covert channels is a vital setup in the analysing the strength of security in a network. Covert Channel is illegitimate channelling over the secured channel and establishes a malicious conversation. The trap-door set in such channels proliferates making covert channel sophisticated to detect their presence in network firewall. This is due to the intricate covert scheme that enables to build robust covert channel over the network. From an attacker's perspective this will ameliorate by placing multiple such trapdoors in different protocols in the rudimentary protocol stack. This leads to a unique scenario of "Hybrid Covert Channel", where different covert channel trapdoors exist at the same instance of time in same layer of protocol stack. For detection agents to detect such event is complicated due to lack of knowledge over the different covert schemes. To improve the knowledge of the detection engine to detect the hybrid covert channel scenario it is required to explore all possible clandestine mediums used in the formation of such channels. This can be explored by different schemes available and their entropy impact on hybrid covert channel. The environment can be composed of resources and subject under at-tack and subject which have initiated the attack (attacker). The paper sets itself an objective to understand the different covert schemes and the attack scenario (modelling) and possibilities of covert mediums along with metric for detection.*
.

## 1. INTRODUCTION

Global internet consists of massive devices connected to it with numerous applications running on it. There is frequent inherent threat of intentional exposure of the confidential and sensitive information over secured channel. Such threats are implemented using **"Covert Channel"** which compromises very important attribute **"Privacy"** of secured channel. Covert channel is defined in different ways based on scenarios of establishment of covert channel and is non-concrete.

**"An enforced, illicit signaling channel that allows a user to surreptitiously contravene the multi-level separation policy and un-observability requirements of the [target of evaluation]."**

DOI : 10.5121/ijnsa.2015.7304   39

This clearly states the policy violation constraint, but does not consider whether the communication channel was envisaged as a communication channel by the system designer. A simple covert channel can be visualizedin [3] where channel comprises of both covert and overt channel in the communication.

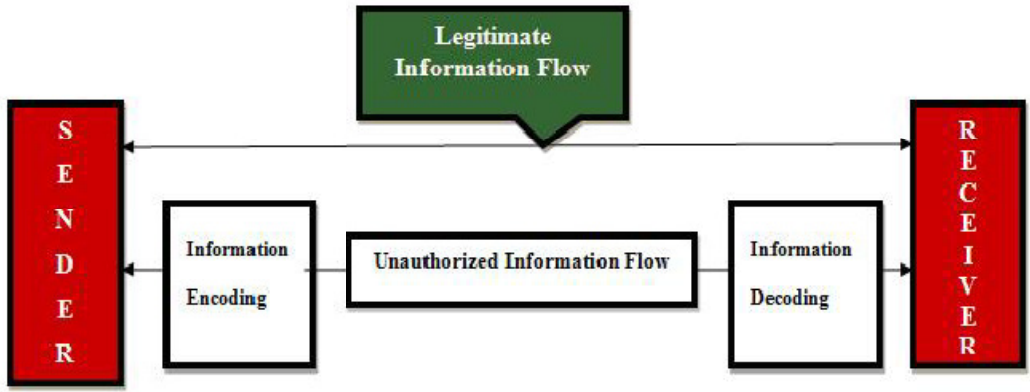

Fig.1. Covert Channel Visualization

Covertchannelinformationexchangeisbasedcovertlanguagespre-negotiated by the covert users and implementation of such languages uses intricate encoding schemes. These schemes may be proliferated into multiple protocols, where each such protocol will be a trapdoor. This makes it complex to detect such clan-destine mediums. SETUP attack [18] makes uses of multi-trapdoor mechanism for ameliorated development of covert channel. A hybrid covert channel scenario may have such multiple trapdoors either in the same layer on in different layers.

Multiple trapdoors can be implemented in the same layer or in different layers.Implementation of the different covert channel variants at the same instance of time tends to behave as a single coherent covert channel. Such channel is termed as **"Hybrid Covert Channel"**. A Hybrid Covert Channel [3] is homogeneous composition of two or more covert channel variants existing at same instance of time. Hybrid covert channel may not have strict composition. It becomes complicated to assess the composition of the Hybrid Covert Channel. An instance of the hybrid covert channel is depicted in [3] and figure 2.

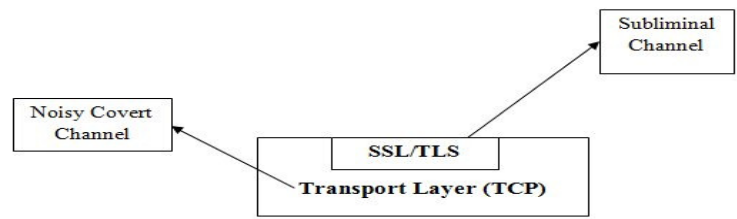

Fig.2. Hybrid Covert Channel in Transport Layer

The covert channel was first introduced in the traditional confinement problem as described in [11]. Extensive work is carried out in devising the detection methods which can be on real-time or forensics [6] based. Scenario based analysis of the covert channel detections [3][7] is performed to understand the detection better. Monitoring the unusual traffic [14] in the network stream is a basis for detection. Modelling the covert timing channel process as Poisson's distribution is also a way to detect such activity. Illegitimate information flows can be tracked through Message Sequence Charts (MSC) [9].This paper employs a statistical protocol based entropy detection [1] to detect hybrid covert channel based on analysis made on packet headers.

## 2. COVERT COMMUNICATION TYPES

In Network communication, covert communication amongst a pair of users can take two forms;
  (a) covert data exchange and
  (b) covert indication

In covert data exchange, covert data is exchanged between the covert users by hiding covert data in rudimentary protocols. This form of covert communication can best be understood with pipeline problem, where there exists two pipes $p_1$ and $p_2$ of diameters $d_1$ and $d_2$ respectively, one inside the other such that $d_2 < d_1$. These pipes are setup between two geographical places for the transportation of crude oil. In Figure 3, the inner pipe $p_2$ of diameter $d_2$ is the covert pipe not known or undocumented in the design and used for smuggling oil. The outer pipe $p_1$ is the legitimate pipe. This type of the covert communication type will not have pre-defined encoding schemes will be simple placement of covert data (trapdoor creation) directly in to the identified clandestine field in the traditional network protocol stack. This channel is called as simple network covert channel.

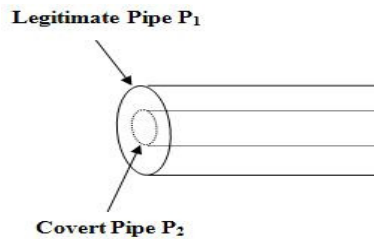

Fig.3. Classical Pipeline Problem

Second form of covert communication is the covert indication. Covert users communicate in a language not known to others. In Figure 4, the covert sender and receiver share an information encoding scheme to leak information. This information encoding scheme as seen from the figure 1 is the language that covert users employ to communicate in a secured legitimate network environment. This sophisticated communication is visible to our detection engine; however decoding the language might be quite difficult in many situations.

The best real time classical example of such communication is Examination Problem. Student X leaks the answers to Student Y for an objective type examination paper in an examination hall in presence of invigilating officer. For each choice in a question, student X makes a gesture that triggers an event to student Y. For instance to communicate choice A to student Y, student X coughs. Same schema holds good in case network communication where covert user X triggers continuous clock events that communicate some form of action to be performed by covert user Y.Some of the other forms of covert indication in network scenario include

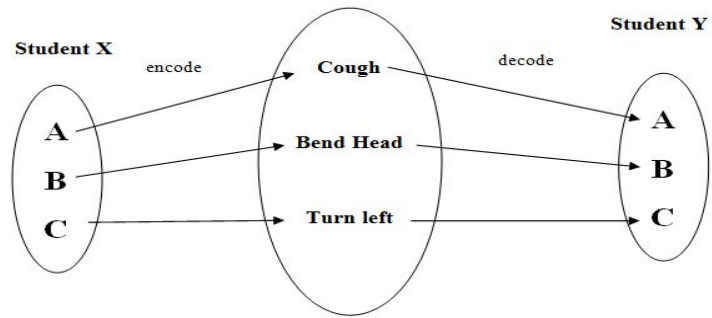

Fig.4. Classical Examination Problem

- Encoding ASCII character set in Sequence number. Decoding the same by applying mathematical operation on sequence number. This can either be in TCP or In IP ID fields.
- Repeated sending of acknowledge packet to an unknown server where the covert receiver is listening to. Receiver has to count the number of time the acknowledge packet was sent to this server. This value can later on mapped to ASCII table for retrieving suitable character.
- Retrieving the packet sorting order numbering in IPSec frames which serves as information to the covert receiver.
- Using logical operators like the XOR with sequence number to get the covert data.

## 3. COVERT CHANNEL VARIANTS

Covert channel are categorized based on different aspects of the overall entities involved in the communication like the shared resources, backdoor/trapdoor placement and parties involved in the communication. The covert channel general classification is given below –

- **Noisy Covert Channel** [14] is a communication channel which has presence of both Overt and covert users.
- **Noiseless Covert Channel** [14] is the communication channel used solely by covert parties.
- **Storage Covert Channel** [14] involves the sender and the receiver either directly or indirectly read or writes in to storage location. The implementation can be on file-lock, R/W in hard disk.
- **Timing Covert Channel** [14] [13] involves the sender signalling the information by modulating the resources in such a way that real response time is observed by the receiver.
- **Simple Network Covert Channel** [14] (SNCC) exists by creating a trapdoor in rudimentary protocols used in network protocol suite.
- **Steganographic Channel** [3] is a means of communication where sender and receiver collude to prevent an observer being able to reliably detect whether communication is happening.
- **Subliminal Channel** [15]- is a covert channel in a cryptographic algorithm, typically proved undetectable.
- **Supraliminal Channel** [12] - A supraliminal channel encodes information in the semantic content of cover data, generating innocent communication in a manner similar to mimic functions.

- **Hybrid Covert Channel** [4] is co-existence of two or more different variants of covert channels existing at same instance of time. The composition of the Hybrid covert channel is difficult to assess from third party which is trying to detect. Mixed composition of covert channel variants behave as single coherent covert channel and is of a greatest threat to the legitimate network environment. For instance noisy covert channel in transport layer with subliminal channel in network layer or application layer.

## 4. ATTACK MODELLING

The attack modelling [4] can be based on different scenarios and placement covert users. Each of these scenarios are designed and built to fulfil certain objectives. Covert users can communicate in direct or encoded format; direct communication is merely placement of covert data over an clandestine medium in the network protocol. Alternatively the covert user can communicate using encoding scheme and that is known only to the covert users.

The intricate design, choosing of clandestine mediums (trapdoors) and encoding scheme paves a way for successful undetectable establishment of covert channel. Detecting such strong covert mediums may be difficult and hence detection metric called covertness index is used. The metric is given below and will be used for assessment in the attack scenarios.

$$\text{Covertness index } (\eta) = \begin{cases} 0 & trapdoor\ is\ detectable \\ \frac{1}{2} & trapdoor\ is\ likely\ detectable \\ 1 & Not\ Detectable \end{cases}$$

This important formation scenarios of covert channels where attack can be devised is given below.

### 4.1 Scenario - 1

The attack scenarios have three entities - Alice, Bob and Eve; Alice and Bob are covert attackers and Eve is legitimate entity/user. The scenario comprises of the combination of covert and legitimate users hence it is scenario of noisy covert channel. The channel established between Bob and Eve is legitimate channel comprising of covert channel and between Alice and Eve is covert channel. Alice and Bob have pre-established channel to communicate the attack information and is mentioned in dotted lines in the figure 5.

While Eve is communicating with Bob over legitimate channel, Alice would extract information over the covert channel. Once when the communication between Bob and Eve is over, Alice would also stop communication with Bob. Further Alice and eve can share the information snatched from Bob's machine. The covert channel implemented between Alice and Bob can have strong trapdoor so as to thwart the detection methods. Such trapdoors can be designed using Hybrid covert channel. Such possible composition can be Subliminal channel in the IPSec and Network Covert Channel in the IP, both at network layer.

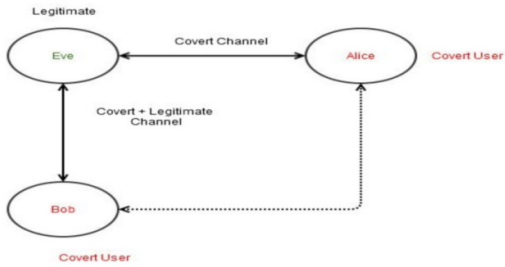

Fig.5. Noise Covert Channel

This combination will prove effective in hop-to-hop routing and can avoid any detections. The covertness index for Network Covert Channel in Network Layer (IPv4)-

$$\eta_{NCC} = \frac{P(T)}{card(U_t)}$$

where P(T) = Probability ofa trapdoor card

(Ut) = Universal set of all possibletrapdoors

$$\eta_{NCC} = \frac{1}{4} = 0.25$$

The covertness index for subliminal channel in IPSec - ESP format

$$\eta_{SCNL} = \frac{number\ of\ rounds\ used\ to\ generate\ S}{max\ number\ of\ rounds\ used\ to\ generate\ r} * trapdoors\ in\ ESP\ format \quad (2)$$

IPSec make use of AES-XCBC-MAC cipher suite and ESP format allow two trapdoors implantation - Sequence Number field and padding. The maximum number of rounds in AES random number generator algorithm is 16. Out of which 5 rounds are used for generating the seed.

$$Covertness\ Index = \frac{5}{16} * \frac{1}{2} = 0.15$$

$$Total\ Covertness\ Index = 0.25 + 0.15 = 0.40$$

As per [7] the trapdoors can be detected under the assumption stated in the hybrid covert channel formation. However this will not be the same if multiple trapdoors are set in each of the protocol headers.

## 4.2 Scenario-2

This scenario is built on the threat model of noiseless covert channel, where the resources and users in sub-network are compromised. This sub-network is connected to other network. The communication from the sub network to all the other networks is built using a Hybrid Covert Channel. This sub network can be similar to bot-net as described in [8].

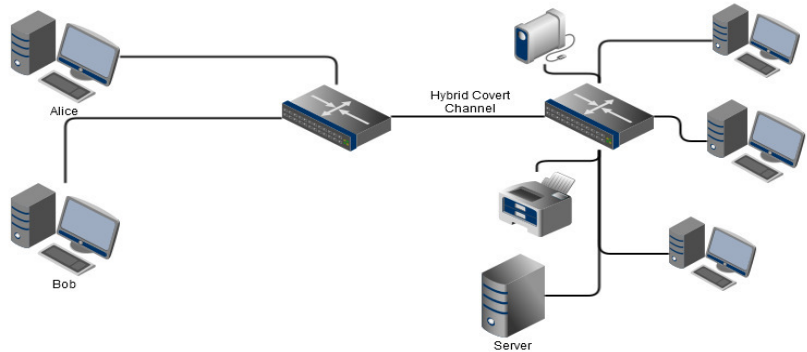

Fig.6. Noiseless Covert Channel with Hybrid Covert Channel

The scenario can have multi-trapdoor or protocol hopped hybrid covert channel [16]. The trapdoor can move from one protocol to another protocol during the hop-to-hop communication or can be combination trapdoors in multiple level in the protocol suite. Hence there can be no particular index.

## 5. COVERT SCHEMES AND THEIR EMBODIMENT

The covert schemes are crucial for conveying the covert data over communication channel in a obscured way. More sophisticated scheme likely not to be retrieved by detection entity. Few samples of covert schemes were discussed in section 2 of this paper and detailed schemes are presented here.

### Scheme 1

The IP ID is field used for identification of the packet and is used for the routing purpose. The covert scheme used for this field is based on following strategy-
- Intentional use of only certain IP ID's while having conversation with Covert receiver.
- Scheme is designed by the covert sender for embedding covert characters in to the IP ID field.
- The Covert receiver applies the scheme used by the sender to retrieve the covert character.

For instance a simple scheme that can be used for this field is extracting the IP ID is by performing modulus operation of the character set size. General notation for this scheme for encoding a character 'c' is

$$E(c) = (R - 1) mod\ n$$

Where $E(c)$ is the encoding function, R is the IP ID value and n is the size of the character set. For an ASCII character set, n = 256

**Example:** If IP ID = 26702 and if the character to be sent is `M' Then $E(M) = 26702 - 1\ mod\ 256$ = `M'

To convey a covert message, the covert sender has select IP ID in such a way as to match with $E(c)$.

**Scheme 2**

Another prominent scheme used is on the sequence number where maximum range is 4,294,967,296 numbers as it is 32 bit field. To communicate covertly under this scheme following strategy is employed-

- Sequence number is multiplied with value of character set and bound is declared with maximum limit.
- The receiver side retrieves the sequence number and then divides it by character set size. The encoding function $E(c)$ is given below-

$$E(c) = (S * n)$$

Where S is the initial sequence number and n is the size of the character set. The decoding function is $D(c')$ is given below –

$$D(c') - S'/n$$

Where $c'$ is the decoded character and $S'$ is the received sequence number.

For instance to send a character `I' covertly over the channel, the sender would have to choose 1235037038 as sequence number and the max. value is derived as 65535 * 256 = 16777216

Therefore the decoded character is $D(c')$ = 1235037038=16777216 = 73, The value 73 when mapped back to ASCII Table is the character `I'.

**Scheme 3**

Another scheme which has tremendous effect on the bandwidth is the modulation of TCP timestamps or use of timing element in the network protocol. TCP timestamps is in the options field of the TCP header which indicates the round trip time of the packets. The TCP process accurately calculates the next retransmission of TCP segment which was failed to be acknowledged. If the character is to be covertly sent using this scheme following strategy is used.

- Get the binary representation of the character and extract bits from the least significant bit.
- Check if the Timestamp least significant bit (LSB) is same as covert bit, if so send the

International Journal of Network Security & Its Applications (IJNSA) Vol.7, No.3, May 2015

   TCP segment.
- Covert receiver will extract the LSB of the timestamp and store the same until it is a byte.

Let $B_c$ be the binary representation of the character `c' and $F_{LSB}(B_c)$ be the encoding function for encoding the covert bits in TCP timestamp.

## 6. ENTROPY BASED COVERT CHANNEL ANALYSIS

The entropy [2] in communication network indicates the number of bits required to encode a character over the channel as stated by Shannon Entropy theory. This is based on the frequency of the characters in given string and the size of the alphabet. The entropy measure also checks for uncertainty of the random variable.

Let A be finite set of characters such that $|S| \geq 1$ and any character `$c' \in A$. A is sequence of symbols which is a string, each of alphabet in string $\in$ A. For instance let **cbbacabbac** be sequence of symbols that needs to be transmitted over network then its sequence of bits represents the coded symbol sequence which may be 101110011011100010. Then the entropy for such scenario is defined as –

$$H(p_i) = \sum_{i=1}^{n} p_i \, log_2 \, p_i$$

where $i \in |S|$ and $|S| > 1$, pi is the probability of the occurrence of symbol 'c' in the string and n gives the length of the string. To transmit a message **"network"** over the communication network, following are the calculated entropy for each alphabet –

The frequency of all the characters in a string with unique symbols will be same, since the word **"network"** has unique symbols the frequency is 0.143. Let X be string for which the entropy is to be calculated, here X may word like network or stream of numbers then

H(X)=[(0.143log20.143) + (0.143log20.143) + (0.143log20.143) +(0.143log20.143) + (0.143log20.143) + (0.143log20.143) + (0.143log20.143)]

H(X)=2:803

It requires 3 bits to represent each symbol in the given string and 21 bits are required to represent the entire string. Further the appropriate line coding technique has to be chosen to represent them in the transmission line. So in general entropy of X where each alphabet is a unique symbol is

$$H(X) = 3 \, x \, |X|$$

In a covert channel scenario, the covert user has to be chosen the message in such a way that the entropy of string should always be less that number of bits available for that field in the protocol header.

i.e., $H(X) < |Maximum\, number\, of\, bits\, in\, that\, field (B_f)|$

47

The IP ID presented in the scheme 1 of this paper has 16 bits in the IP header, so to send X the minimum of 21 bits are required. Hence capacity of the covert channel is

$$C_c = log_2(1 + \tfrac{16}{21}) = 0.25$$

The covert channel occupies 25% of total IP header space. Multiple trapdoors (t) [5] [4] in IP header or protocol header simply doubles the covert channel capacity. However the entropy to channel capacity ratio will be low thus making it robust ie.,

$$\frac{0.25 x 2}{2.803} = 0.1\ddot{7}$$

This makes the detection of covert bits much difficult as the detection systems needs to scan more fields for analysis.

In general,

$$\frac{C_c * t}{H(X)} < H(X)$$

for robust covert channel construction where η [7]the covertness index for such multi-trapdoor covert channel will be greater than 0.5. The multiple trapdoors through a protocol or set of protocols is actually setting up of multiple covert channels in the communication network. The entropy for such scenarios is dispersed across multiple making it difficult to understand the scheme. Also in the scenario of multi-trapdoors covert channel behaves like a single coherent hybrid covert channel where the effect of the entropy is doubled. The below results shown in the figure 7 and figure 8 shows the accurate expected behaviour discussed in this paper -

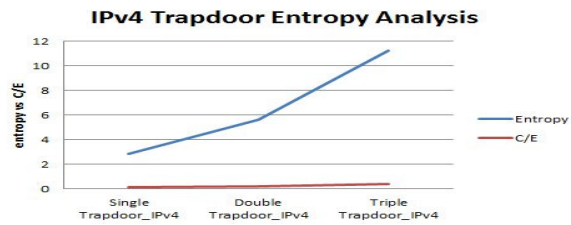

Fig.7. IP Entropy analysis

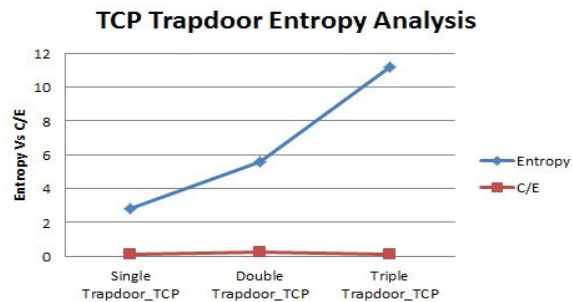

Fig.8. TCP Entropy Analysis

The results indicate the multiple trapdoors used in hybrid covert channel yields to a higher entropy value and low channel to entropy ratio (C/E). The constant C/E ratio also indicates the consistent usage of protocol header for constructing multi-trapdoor based hybrid covert channel. This implies that the covert schemes used in Hybrid covert channel is difficult detect in secured communication.

## 7. RESULTS AND DISCUSSIONS

The number of trapdoors implemented in a protocol cannot be all the fields vulnerable in that protocol. i.e.,

$$\delta = T_m - T_s \text{ and } T_s < T_m$$

where $T_m$ is the max number of trapdoors possible in that protocol $T_s$ is the no. of Trapdoors set.

The analysis of the trapdoor setting is performed on protocols like IPv4, TCP and IPSec, SSL/TLS. The trapdoor setting in the rudimentary network protocols like the IPv4 and TCP is merely based on placing the covert data in any of its header fields. The table 1 shows effect of varying the number of trapdoors in IPv4 protocol.

Table 1. Multi-Trapdoor Analysis of IPv4

| Sl.No. | TrapdoorName | No.of Trapdoors | No. of Trapdoor | Algorithm | Covertness Index | Entropy | C/E |
|---|---|---|---|---|---|---|---|
| 1 | NetworkCovert Channel-IPv4-Single | 4 | 1 | NIL | 0.25 | 2.803 | 0.089 |
| 2 | NetworkCovert Channel-IPv4-dual | 4 | 2 | NIL | 0.5 | 5.606 | 0.17 |
| 3 | NetworkCovert Channel-IPv4-triple | 4 | 3 | NIL | 0.75 | 11.21 | 0.358 |

The graph of Trapdoors Vs the Covertness Index is show in the figure 9 where increase in the number of the trapdoors in IPv4 increases the difficulty in detecting the covert channel. The trapdoor setting in IPSec using subliminal channel is slightly complex to understand. However the ESP format provides two fields to convey the covert bits in the protocol header. The remaining data is sent over the ESP algorithm during the time of the key generation for encryption using AES algorithm. The residual bits in used in random number generation or used in the round box of the AES and this is depicted on row 2 of the table 2. Hence the covertness index is 0.15 * equation 2 which is 0.47. This will not change any further as there is limited scope for subliminal channel development in IPSec -ESP format.

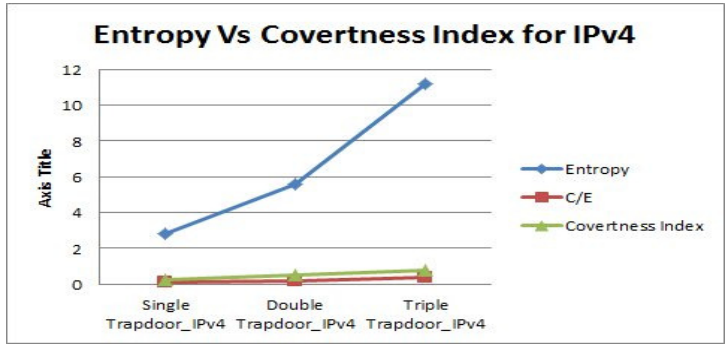

Fig.9. Entropy Vs Covertness Index in IPv4

Table 2. Multi-Trapdoor Analysis of Subliminal Channel in IPSec

| Sl.No. | TrapdoorName | No.of Trapdoors | No. of Trapdoorsu | Algorithm | CovertnessIndex | Entropy | C/E |
|---|---|---|---|---|---|---|---|
| 1 | SubliminalChannel-IPSecESP-1 | 2 | 1 | AES-XCBC-MAC | 0.15 | 2.803 | 0.14 |
| 2 | SubliminalChannel-IPSecESP-2 | - | - | AES-XCBC-MAC | 0.47 | 4.78 | 0.35 |
| 3 | SubliminalChannel-IPSecESP-3 | - | - | AES-XCBC-MAC | 0.47 | 5.21 | 0.35 |

The graph of Trapdoors Vs the Covertness Index is show in the figure 10 where increase in the number of the trapdoors in IPSec ESP makes covertness index constant. The trapdoors in TCP based protocol is simple and provides seven fields for placing the covert data. The table 3 depicts the changing trapdoor that has an effect on the covertness index. When more number of the trapdoors are involved it is difficult to detect the composition of the covert channel. The figure 11 shows change in the trapdoor count that has an effect in the detection. However the changes in the covertness index can be minimal. The trapdoor setting in the subliminal channel in SSL/TLS is based on the algorithm used in its cipher suite. This is purely called as random oracle channel. However to increase the complexity of the subliminal to thwart detection the randomization of bits is feasible in chosen prime number. This forms Newton Subliminal Channel. The covertness index for such channels is discussed in the table 4

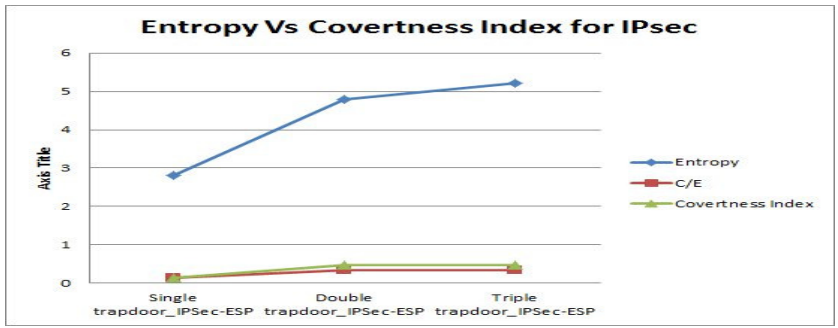

Fig.10. Entropy Vs Covertness Index in IPSec based subliminal channel

Table 3. Multi-Trapdoor Analysis of Network covert channel in TCP

| Sl.No. | TrapdoorName | No.of Trapdoors | No. of Trapdoorsu | Algorithm | Covertness Index | Entropy | C/E |
|---|---|---|---|---|---|---|---|
| 1 | NetworkCovertChannel-TCP-1 | 7 | 1 | NIL | 0.142 | 2.803 | 0.14 |
| 2 | NetworkCovertChannel-TCP-2 | 7 | 2 | NIL | 0.28 | 5.606 | 0.28 |
| 3 | NetworkCovertChannel-TCP-3 | 7 | 3 | NIL | 0.42 | 11.21 | 0.14 |

The graph of covertness index Vs the trapdoor in the subliminal channel is shown in the figure 12. The higher entropy value for the some of the formation indicates that the detection engine [10] is able to detect the activity and this give clear indication of the higher detection rates. Hybrid Covert channel is not feasible for the combinations of the Network covert channel in TCP and IPv4 as this become easily detectable combination.

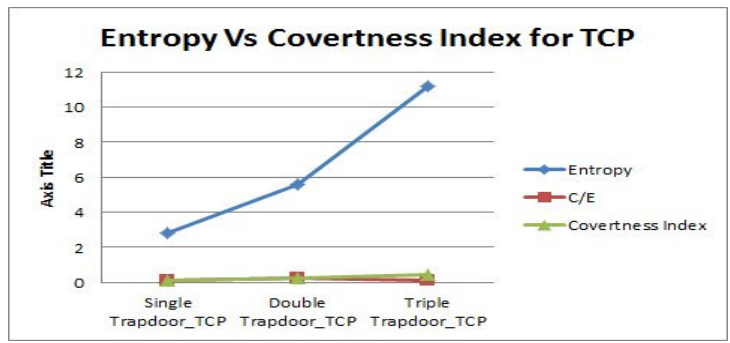

Fig.11. Entropy Vs Covertness Index in Covert Channel based on TCP

Table4.Multi-TrapdoorAnalysisofSubliminalChannelinSSL/TLS

| Sl.No. | TrapdoorName | No.of Trapdoors | No. of Trapdoorsu | Algorithm | Covertness Index | Entropy | C/E |
|---|---|---|---|---|---|---|---|
| 1 | SubliminalChannel(Oracle)-SSL/TLS-1 | - | - | SSLCipherSuite | 0.25 | 2.803 | 0.14 |
| 2. | SubliminalChannel(Oracle)-SSL/TLS-2 | - | - | SSLCipherSuite | 0.58 | 3.67 | 0.35 |
| 3 | SubliminalChannel(Oracle)-SSL/TLS-3 | - | - | SSLCipherSuite | 0.58 | 3.67 | 0.35 |

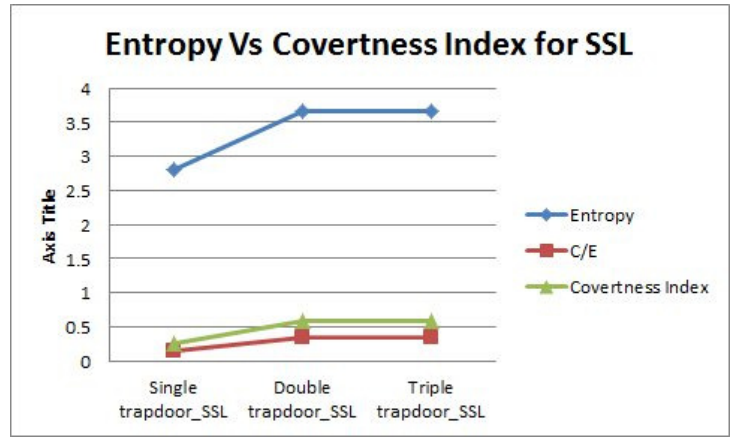

Fig.12. Covertness Index for Subliminal Channel based on SSL/TLS

## 8. CONCLUSION

Covert schemes are difficult to understand from third party entity as they obscure the content taken in protocol header. This provides an opportunity for embedding any data which may even be malware code. Entropy based analysis gives the actual number of bits used to represent the covert symbol in a protocol. This gives clearly metric to understand the covert channel schemes in a better way. It is unacceptable to have malicious conversation of the network even in presence of administrator. It is inference this experiment that the hybrid covert channel has high degree of entropy which makes it difficult to detect. It is required to concentrate on stronger detection principle to detect such events.

## ACKNOWLEDGEMENT

AnjanKoundinya thanks Late Dr. V.K Ananthashayana, Erstwhile Head, De-partment of Computer Science and Engineering, M.S.Ramaiah Institute of Tech-nology, Bangalore, for igniting the passion for research.

**AUTHOR'S**

**AnjanK** has received his B.E degree from Visveswariah Technological University,Belgaum,India in 2007 And his master degre from Department of Computer Science and Engineering, M.S.RamaiahInstitute of Technology ,Bangalore, India.He has been awarded Best Performer PG 2010 for his academic excellence.His area so fresearch includes NetworkSecurityandCryptography,Agile Software Engineering.He ispursuing Ph.D in Computer Science and Engineeing fromVTU,Belgaum. He is currently working as Assistant Professorin Dept.of Computer Science and Engineering, RV College of Engineering, Bengaluru, India. 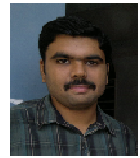

**SrinathNK** has his M.E degree in Systems Engineering and Operations Research from Roorkee University, in 1986 and PhD degree from Avinash Lingum University,India in 2009.His areas of research interests include Operations Research, Parallel and Distributed Computing, DBMS ,Microprocessor. His isworking as Professor and Dean PG, Dept of  Computer Science and Engineering,RVCollege of Engineering. 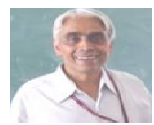

**JibiAbraham** has received her M.S degree in Software Systems from BITS,Rajasthan,India in 199 and PhD degree from Visveswariah Technological University , Belgaum , India in 2008 in the area of Network Security.He rarea so fresearch interests include Network routing algorithms ,Cryptography ,Network Security of Wireless Sensor Networks and Algorithms Design.She is working as Professor and Head in Dept. of CEIT, College of Engineering Pune. 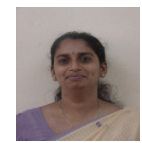